\documentclass[11pt]{article}

\usepackage{abstract}

\usepackage[a4paper]{geometry}
\usepackage{blindtext}

\usepackage{diagbox}
\usepackage{todonotes}
\usepackage{threeparttable}

\usepackage[section]{placeins}
\usepackage{hyperref}
\hypersetup{
  colorlinks=true,
  linkcolor=blue,
  citecolor=blue,
  urlcolor=blue}
\usepackage{color}
\usepackage{xcolor}

\usepackage{amsthm,amsmath}

\usepackage{longtable,booktabs}
\usepackage{threeparttable}
\usepackage[labelsep=period]{caption}
\usepackage[style=authoryear-comp, backend=biber, uniquename=false, maxnames=9, maxcitenames=2, maxbibnames=99, natbib=true, bibencoding=utf8, dashed=false,terseinits=true, giveninits=true, uniquelist=false, doi=false, isbn=false, url=true, sortcites=false, date=year,safeinputenc]{biblatex}
\renewbibmacro{in:}{}
\DeclareNameAlias{sortname}{family-given}

\DeclareFieldFormat{url}{\texttt{\url{#1}}}
\DeclareFieldFormat[article]{pages}{#1}
\DeclareFieldFormat[inproceedings]{pages}{\lowercase{pp.}#1}
\DeclareFieldFormat[incollection]{pages}{\lowercase{pp.}#1}
\DeclareFieldFormat[article]{volume}{\mkbibbold{#1}}
\DeclareFieldFormat[article]{number}{\mkbibparens{#1}}
\DeclareFieldFormat[article]{title}{\MakeCapital{#1}}
\DeclareFieldFormat[inproceedings]{title}{#1}
\DeclareFieldFormat{shorthandwidth}{#1}
\usepackage{verbatim}
\usepackage{indentfirst}

\usepackage{bm}

\let\proglang=\textsf

\title{Bayesian forecast combination using time-varying features}

\begin{document}

\author{Li Li \footnote{School of Economics and Management, Beihang University, Beijing
100191, China. Email: \texttt{by1908007@buaa.edu.cn}, ORCID:
\url{https://orcid.org/0000-0002-7922-1281}. }, Yanfei Kang\footnote{School of
Economics and Management, Beihang University, Beijing 100191, China. Email:
\texttt{yanfeikang@buaa.edu.cn}, ORCID:
\url{https://orcid.org/0000-0001-8769-6650}. }, and Feng Li\footnote{ School of
Statistics and Mathematics, Central University of Finance and Economics, Beijing
102206, China.  Email: \texttt{feng.li@cufe.edu.cn}, ORCID:
\url{https://orcid.org/0000-0002-4248-9778}, Corresponding author. }}

\date{}
\maketitle

\begin{abstract}
  In this work, we propose a novel framework for density forecast combination by constructing time-varying weights based on time-varying features. Our framework estimates weights in the forecast combination via Bayesian log predictive scores, in which the optimal forecast combination is determined by time series features from historical information. In particular, we use an automatic Bayesian variable selection method to identify the importance of different features. To this end, our approach has better interpretability compared to other black-box forecasting combination schemes. We apply our framework to stock market data and M3 competition data. Based on our structure, a simple maximum-a-posteriori scheme outperforms benchmark methods, and Bayesian variable selection can further enhance the accuracy for both point forecasts and density forecasts.

\end{abstract}

\clearpage
\section{Introduction}
\label{sec:introduction}

Achieving a robust and accurate forecast is a central focus in finance and econometrics. Forecast combination has been adopted as an essential enhancement tool for improving time series forecasting performance during recent decades \citep{Kolassa2011Combining,Bergmeir2016Bagging,garratt2019real}, due to its ability to reduce the risk of selecting a single or inappropriate model and lead to more accurate and robust forecasts \citep{Jose2008Simple}. See \citet{wang2022forecast} for a comprehensive survey in this area. It is evident that merely tackling the model uncertainty can deliver most of the performance benefits despite that the overall forecasting uncertainty is affected by the model, data and parameter uncertainty \citep{Petropoulos2018Exploring}.  Forecast combination is such an instrument for reducing the overall forecasting uncertainty with the focus on finding optimal weights of different forecasting models in the combination \citep{monteromanso2020fforma,kang2020forecast,kang2019gratis}.

Nevertheless, most existing forecasting combination approaches have been limited to point forecasts, which prevent their extension to decision making problems, especially in domains like economics and business planning.  In finance and economics, investigators rely on the complete insights of the uncertainty proved by the density forecasts \citep{Kascha2010Combining}. See \citet{Tay2000Density} and \citet{timmermann2006forecast} for a review on density forecasts in finance and economics.

Primary research already shows that simply applying a similar strategy of point forecast combination to density combination could generally elevate the forecasting performance compared with choosing a particular model \citep{Kascha2010Combining,liu2009forecasting}.  Thereafter, density forecast combinations have attracted broad attention in recent years \citep{ciccarelli2010forecast,opschoor2017combining}, mainly focusing on how to weight different forecast densities and update weights over time \citep{aastveit2018evolution}.

\citet{Wallis2005Combining} started this line of work and proposed a finite mixture distribution to combine density forecasts. Then \citet{Hall2007Combining} devised a weighted linear combination by minimizing the distance between forecasting and the true-but-unknown densities based on the logarithmic scoring rule.  \citet{Pauwels2016note} structured a series of simulation experiments to examine the properties of the optimization problem in \citet{Hall2007Combining}.  \citet{Kascha2010Combining} considered different combining and weighting schemes for inflation density forecasts evaluated by the average log score.  \citet{jore2010combining} developed a combination strategy for autoregressive models based on log-score recursive weights. \citet{Aastveit2014Nowcasting} proved that the combined density scheme of \citet{jore2010combining} performed better and more robustly than component models in both log scores and calibration tests. A noticeable approach in the literature is the ``optimal prediction pools'' (shorthand for ``OP'' hereafter) proposed by \citet{geweke2011optimal}, in which they used an instructive linear pool to obtain the optimally weighted density combination under scoring criteria. They utilized the historical performance of forecasting models in the pool to determine the weights by maximizing log predictive scores.

Although these combined forecasting methods improve the accuracy compared to the single model, there are still some apparent disadvantages. The properties of forecast performance from different models may change over time, and the result is that the combining method with constant weights may not be the optimal scheme.  We call this ``forecast combination uncertainty''. To cope with the challenge in density forecasting, new combination methods with time-varying weights are also studied. For example, \citet{Waggoner2012Confronting} explored the regime-dependent weights and time-varying importance of two macroeconomic models. \citet{casarin2013parallel} learned time-varying combination weights from past forecasting performance and other mechanisms.  \citet{Kapetanios2015Generalised} put forward the ``generalized pools'' (shorthand for ``GP'' hereafter), extending the OP method \citep{geweke2011optimal} by a more general scheme for combination weights. \citet{Kapetanios2015Generalised} utilized piecewise linear weight functions to make the weights depend on the regions of the distribution and proved that GP produced more accurate forecasts compared with optimal combinations with fixed weights \citep{Hall2007Combining,geweke2011optimal}. ``Dynamic pools'' \citep{Del2016Dynamic} were then provided, relying on a sequence of time-varying weights for the combination of two Dynamic Stochastic General Equilibrium (DSGE) models. Recently, \citet{mcalinn2019dynamic} provided a Bayesian Predictive Synthesis (BPS) framework, encompassing several existing forecast density combination methods in \citet{geweke2011optimal} and \citet{Kapetanios2015Generalised}.

The aforementioned variants of density forecast combination mainly focus on forecasting model uncertainty, neglecting the uncertainty or characteristic of the time series itself. In addition, those methods often lack interpretability. To be specific, they directly obtain the optimal combination weights without explaining what features of the time series affect the weights.  Feature-based time series forecasting has received remarkable applications over the years. \citet{wang2009rule} derived recommendation rules by learning the relationship between time series features and the suitability of forecasting methods. \citet{Petropoulos2014Horses} studied the influence of seven time series features to forecast accuracy and provided helpful conclusions for method selection. \citet{talagala2022fformpp} developed a random forest classifier to select the best forecasting model based on 42 time series features under the meta-learning framework. Then \citet{monteromanso2020fforma} further utilized the 42 features to select weights of each forecasting model and proposed a framework called FFORMA (Feature-based FORecast Model Averaging).

However, the recently developed, especially the machine learning based forecast combination methods are usually black boxes. The internal logic of the combination weights is hard to explain due to the complexity of algorithms.
For instance, FFORMA \citep{monteromanso2020fforma}, utilized 42 expert selected features to determine combination weights with an XGBoost algorithm, ranking second place in the M4 competition. But the relation between the features and weights can not be interpreted due to the ``black-box'' learning algorithm. In this paper, we study this problem from an orthogonal perspective to the existing literature, that is, to explain time-varying weights by time-varying features of time series. Furthermore, our method handles the forecast uncertainties from different aspects. First, the combination of different models reduces model uncertainty. Second, the combination with time-varying weights can deal with forecast combination uncertainty. Third, time series features are applied to capture data uncertainty. We define the time-varying weights by a softmax transform of a linear function of time series features and redefine the log predictive score function \citep{geweke2011optimal}. The main extension of our method in the scoring function is that the weights are determined by features and vary over time. Then the optimal weights are obtained through maximizing the historical log predictive scores in the pool, as in \citet{Hall2007Combining}, \citet{geweke2011optimal}, \citet{Kapetanios2015Generalised} and \citet{Del2016Dynamic}. We estimate unknown parameters in the weights by the maximum-a-posteriori (MAP) method, considering the prior knowledge.

Based on the time-varying weighting scheme, a challenge is to choose relevant features to match the forecast combination and interpret the importance of different features.  Choosing features only based on some intuition or expertise may lead to feature selection bias, especially when forecasters' information is inadequate. Nonetheless, vast time series features are proposed in the literature and software recently. It is impossible for practitioners to always pick the right set of features. Putting all possible features into the combination model not only scales up the computational difficulty but also reduces the variable selection efficiency. Because it is well-known that straightforward Bayesian variable selection does not perform well in a very large variable set.

Inspired by the statistical screening methods for variable selection, we introduce an initial screening process to determine some candidate features from a larger feature set and use the ReliefF algorithm \citep{kononenko1994estimating} to pick out a subset of features that shows differences in forecasting performance for different models.  Then we introduce an automatic Bayesian variable selection method to weight the contribution of selected features.

There are five principal advantages of the proposed framework: (1) our approach is more comprehensible than black-box forecasting combinations as not only interpreting which features determine the combination weights but also identifying the importance of different features; (2) the combination weights can vary over time based on time-varying features and handle diversiform uncertainties from the model, forecast combination and data; (3) a complete Bayesian framework is formed and prior information in the combination are taken into consideration; (4) the framework is computationally efficient because we can calculate some steps in the offline phase and our algorithm is easy to parallel with large time series sets; and (5) our framework can produce both point forecasts and density forecasts in one step or multiple steps, which makes it more flexible than OP \citep{geweke2011optimal} and GP \citep{Kapetanios2015Generalised} methods.

The rest of the paper is organized as follows. Section \ref{sec:forecasts-evaluation} introduces the feature-based Bayesian forecasting evaluation metric. Section \ref{sec:forecasts-inference} proposes a general framework for forecast combination by establishing the Bayesian inference scheme using the feature-based log predictive score. In Section \ref{sec:stockdata}, we apply the proposed framework to a Standard and Poors (S\&P) 500 index data. An extensive collection of the M3 data set is used in Section \ref{sec:m3data} to further demonstrate the superiority of our framework. Section \ref{sec:discussion} provides our discussions and Section \ref{sec:conclusions} concludes the paper.

\section{Bayesian forecasting evaluation with features}
\label{sec:forecasts-evaluation}

With a full density forecasting approach, forecasting performance is then can be measured by a joint predictive probability \citep{geweke2001bayesian,geweke2010comparing}. For a target series $Y_{T}=\left\{ y_{\text{1}},y_{2},\cdots ,y_{T} \right\}$, the conditional probability density for $y_t$ of a single forecasting model $M$ is
\begin{equation}
  \label{eq:p}
    p(y_{t}|Y_{t- 1}, M)  = \int p\left( y_{t},{\theta }_{M}|Y_{t - 1}, M\right) d{\theta}_{M}
    = \int p\left( y_{t }|{\theta }_{M}, Y_{t - 1}, M\right) p\left({\theta }_{M}| Y_{t - 1}, M\right) d{\theta}_{M}
  \end{equation}
  and $p\left({\theta }_{M}| Y_{t - 1}, M\right)$ is the posterior of unknown parameters in model $M$ given the historical data.

The logarithmic of joint predictive probability is known as the log score (LS)
\begin{equation}
  \label{eq:LS}
    LS \left( Y_{T}, M\right)=\sum_{t = 1}^T \log
    p(y_{t}|Y_{t- 1}, M),
\end{equation}
which measures the out-of-sample forecasting performance and a larger value of LS indicates that the forecasting is more accurate. The LS becomes a popular tool to measure the accuracy of density forecasts \citep{Gneiting2007Strictly,gaglianone2014constructing,mitchell2011evaluating}.

As mentioned earlier, the performance of combined forecasting methods is often better than that of an individual model on account of reducing model uncertainty. We extend the predictive probability for a pool of models with a linear form of many individual predictive densities as
\begin{equation}
  \label{eq:pred_mixture}
    p\left( y_{t}| Y_{t - 1}, \mathcal{M}\right)=\sum\limits_{i=1}^{m}{{{w}_{i}} p\left( y_{t}| Y_{t - 1}, M_i\right)},
\end{equation}
where $\mathcal{M} = \left\{M_1, M_2, ..., M_m \right\}$ and ${w}_{i}$ is the weight of model $M_i$ in forecast combination.

In recent years, a large number of scholars have proposed effective forecasting selection or combination methods based on time series features through rule-induction \citep{arinze1994selecting}, meta-learning \citep{monteromanso2020fforma,wang2009rule}, random forecast classifier \citep{talagala2022fformpp} and so on. We firstly use time series features to construct time-varying weights in the forecast combination by a variant of softmax function as
\begin{equation}
  \label{eq:wit}
    {w}_{i, t}=\frac{\exp \left\{ x_{t}'{{\beta }_{i}} \right\}}{1+\sum\nolimits_{i=1}^{m-1}{\exp \left\{ {x}_{t}'{{\beta }_{i}} \right\}}},i=1,2,\cdots ,m-1,
\end{equation}
where ${{w}_{i}}$ is the feature-dependent weight of the $i$-th model, $x_{t}$ is a vector of features calculated from historical data available at time $t-1$ and $\beta_i$ is the coefficient vector of features, only related to the $i$-th model (invariant with time). The non-global time series features here are calculated with a moving window to make them time-varying.  Equation (\ref{eq:wit}) has the same functionality of the standard softmax function but avoids the identification issue in the mixture contexts \citep{fruehwirth-schnatter2006finite}. The log predictive score for a pool of models is redefined as
\begin{equation}
  \label{eq:logpred_likelihood}
    LS( Y_{T}, \mathcal{M})=\sum_{t = 1}^T \log\left[ \sum\limits_{i=1}^{m}{{{w}_{i, t}} p\left( y_{t}| Y_{t - 1}, M_i\right)}\right].
\end{equation}

It is worth mentioning that the evaluation function of the weighted linear pool proposed by \citet{geweke2011optimal} is a special case of Equation (\ref{eq:logpred_likelihood}) (when the coefficients of features are all zero). The weights in OP \citep{geweke2011optimal} are obtained based on the historical performance of the pooled models by maximizing the log predictive score. However, the weights of OP \citep{geweke2011optimal} are constant over time, leading to the limitation of capturing the combination variation.

\section{General framework for Bayesian forecast combination}
\label{sec:forecasts-inference}

\subsection{The Bayesian scheme}

To carry out a full Bayesian inference scheme for the coefficients of features in Equation (\ref{eq:wit}), we present the posterior of the feature coefficients based on the predictive likelihood as
\begin{equation}
  \label{eq:febama}
    p\left( \beta |{{Y}_{T}}, {{X}_{T}}, \mathcal{M}\right)\propto \prod\limits_{t=1}^{T}{p(y_{t}|Y_{t- 1}, {{X}_{t}}, \beta, \mathcal{M})} p\left( \beta  \right),
\end{equation}
where ${{X}_{T}}$ is the time series feature matrix of historical data, $\beta$ is the unknown parameter matrix and $p\left( \beta \right)$ is the prior.  Then the log posterior is connected with the log score in Equation (\ref{eq:logpred_likelihood}) as
\begin{equation}
  \label{eq:febama_log}
    \log p\left( \beta |{{Y}_{T}}, {{X}_{T}}, \mathcal{M}\right) = \mathrm{constant} + \sum_{t = 1}^T\log\left[ \sum\limits_{i=1}^{m}{{{w}_{i, t}} p\left( y_{t}| Y_{t - 1}, M_i\right)}\right] + \log p\left( \beta  \right).
\end{equation}
In this study, we call the Bayesian approach formulated by Equation \eqref{eq:febama_log} FEBAMA (FEature-based BAyesian forecasting Model Averaging).

\begin{figure}
  \centering
  \includegraphics[width=\textwidth]{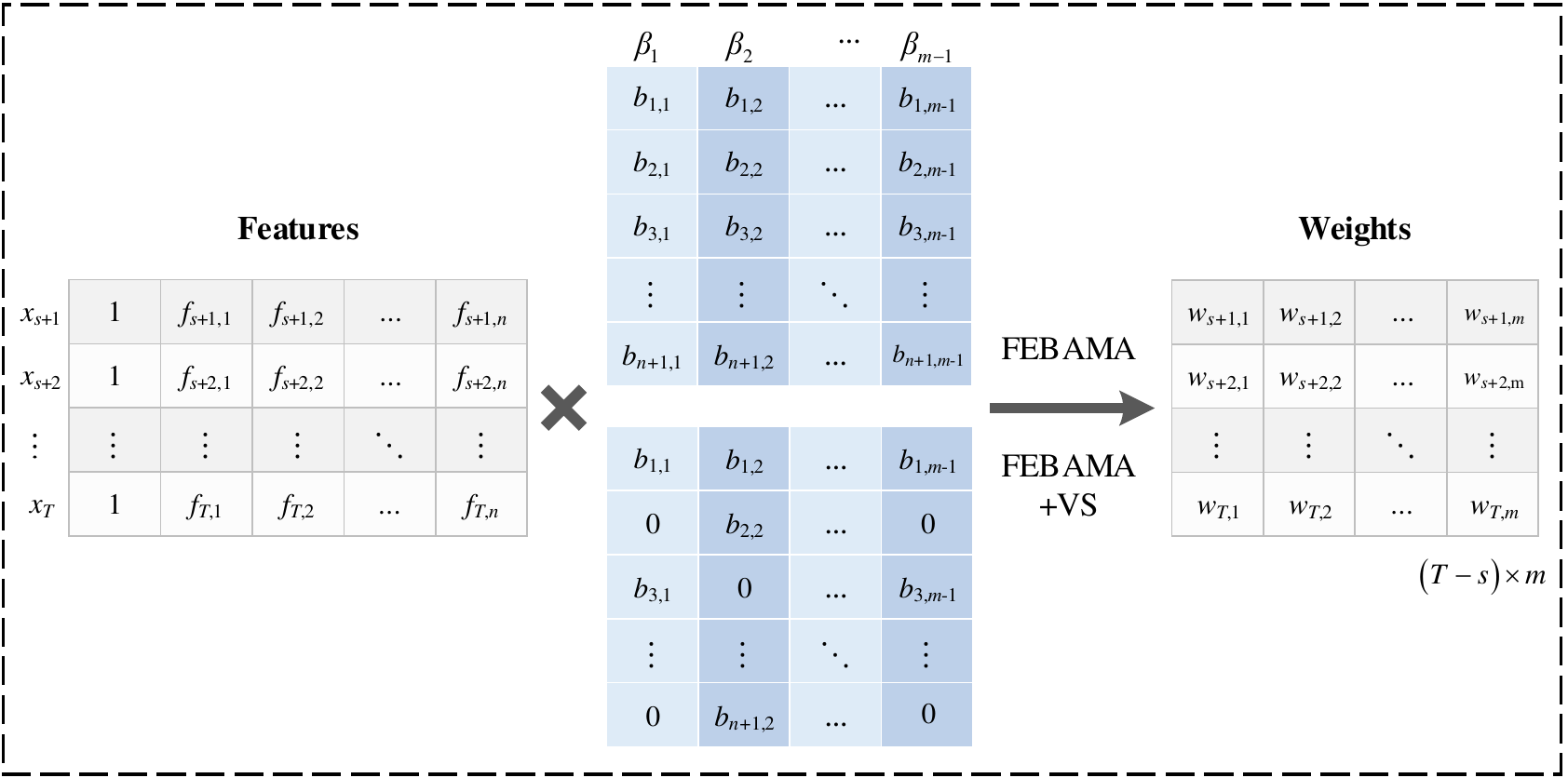}
  \caption{The computational process for obtaining time-varying weights in FEBAMA and FEBAMA+VS. For the target series $Y_{T}$, we arrange the minimum length to be $s$ to generate time series features.  The upper parameter matrix is for FEBAMA. The lower parameter matrix within some zero values is for FEBAMA+VS on the condition of variable selection indicators.}
  \label{weights}
\end{figure}

With the above setup, a Bayesian variable selection method can be directly applied to select important time series features, which we denote as ``FEBAMA+VS''. Let ${\mathcal{I}_{i}}$ be the variable selection indicator vector for the model $M_i$, and ${\beta }_{\mathcal{I}_i}$ is the corresponding coefficient vector of features. So the expression for the combination weights is changed to
\begin{equation}
  \label{eq:wti_vs}
    {w}_{i, t}=\frac{\exp \left\{ x_{t}'{{\beta}_{\mathcal{I}_i}} \right\}}{1+\sum\nolimits_{i=1}^{m-1}{\exp \left\{ {x}_{t}'{{\beta }_{\mathcal{I}_i}} \right\}}},i=1,2,\cdots ,m-1,
\end{equation}
and the joint posterior for both the coefficients and variable selection indicators is
\begin{equation}
  \label{eq:febama_full}
    p\left( \beta, \mathcal{I} |{{Y}_{T}}, {{X}_{T}}, \mathcal{M}\right)\propto \prod\limits_{t=1}^{T}{p(y_{t}|Y_{t- 1}, {{X}_{t}}, \beta, \mathcal{I},\mathcal{M})} p\left( \beta, \mathcal{I}\right),
\end{equation}
where $p\left( \beta, \mathcal{I} \right)$ is the joint prior.  Figure \ref{weights} shows the structure of time-varying weights in our framework. The intercept term is always included in the linear function of features. As shown in the lower parameter matrix in Figure \ref{weights}, when the variable selection indicator is $0$, the corresponding feature is not selected for the corresponding model.

The main contribution of our Bayesian scheme is twofold.  First, our approach connects the Bayesian forecasting evaluation metric, log predictive score, with time-varying features to allow for an interpretable time-varying weighting scheme. The time-varying weights can be learned from historical data. Second, compared to the previous methods, such as OP \citep{geweke2011optimal} and GP \citep{Kapetanios2015Generalised} directly optimizing the score function, we provide a full Bayesian setup to make the forecast combination process a standard Bayesian inference procedure. Our Bayesian scheme could incorporate experts' knowledge for the combining strategy and use an automatic Bayesian variable selection to detect the most important features in the combination weights.

\subsection{Inference procedure}

In the inference procedure, our purpose is to obtain the time-varying weights conditional on the optimal coefficients. We set the priors of indicator and coefficient vectors in Table \ref{tab: priors}, which are homologous with the non-informative priors \citep{gelman2013bayesian} in the Bayesian setting.  Then, we take the MAP estimation with the standard BFGS algorithm \citep{byrd1995limited}. The optimal coefficients can be obtained by maximizing the log posterior, which is consistent with maximizing the LS based on sufficient historical data. Furthermore, the proposed FEBAMA based on time-varying features can achieve a larger LS than OP \citep{geweke2011optimal}, which is demonstrated in the following experiments in sections \ref{sec:stockdata} and \ref{sec:m3data}. It is worth mentioning that the ``optimal weights'' in  \citet{geweke2011optimal} is optimal under the condition of constant combination weights. There is no guarantee that those weights are still optimal in the time varying combinations, which implies our case. Our approach falls back to the ``optimal weights'' in  \citet{geweke2011optimal} when all coefficients are zero.

\begin{table}
  \centering
  \caption{Parameter settings in the priors of  indicator vectors and  coefficient
    vectors.}
  \label{tab: priors}
  \resizebox{\textwidth}{!}{
    \begin{tabular}{lp{5cm}p{8cm}}
      \toprule
      Parameter                  & Description                                                                  & Prior                                                                                     \\
      \midrule
      ${{\beta }_{i}}$           & Coefficient vectors of features for the model $M_i$.                         & Each element follows an independent normal distribution $N(0,{\sigma}^{2})$.              \\
      $\mathcal{I}_i$.           & Indicator vectors of features for the model $M_i$.                           & Each element follows an independent $Bernoulli(p)$ distribution where $p\sim Beta(1, 1)$. \\
      $\beta_{\mathcal{I}_{i}}$. & Nonzero coefficient vectors of features on the condition of $\mathcal{I}_i$. & Each element follows a conditional normal distribution.                                   \\
      \bottomrule
    \end{tabular}
  }
\end{table}

If variable selection is considered simultaneously in our framework, we utilize the Gibbs sampler to perform feature selection over all forecasting models. Therefore, multiple MAP estimations are required. Within each model, a randomly proposed variable selection indicator $\mathcal{I}^{\mathrm{propose}}$ based on the current variable selection indicator $\mathcal{I}^{\mathrm{current}}$ is accepted with the Metropolis acceptance rate
\begin{align}
  \alpha = \mathrm{min} \left( 1,  \frac{p\left(\mathcal{I}^{\mathrm{propose}}, {\beta}_{\mathcal{I}^{\mathrm{propose}},} | {{Y}_{T}}, {{X}_{T}}\right)}{p\left(\mathcal{I}^{\mathrm{current}}, {\beta}_{\mathcal{I}^{\mathrm{current}},} | {{Y}_{T}}, {{X}_{T}}\right)} \right).
\end{align}
The proposed FEBAMA+VS can output the frequencies that different features are selected based on all the accepted variable selection indicators. The frequency quantifies the contribution of selected features to the combination weights.

\subsection{Training and forecasting}

\begin{figure}
  \begin{center}
    \begin{tabular}{c}
      \includegraphics{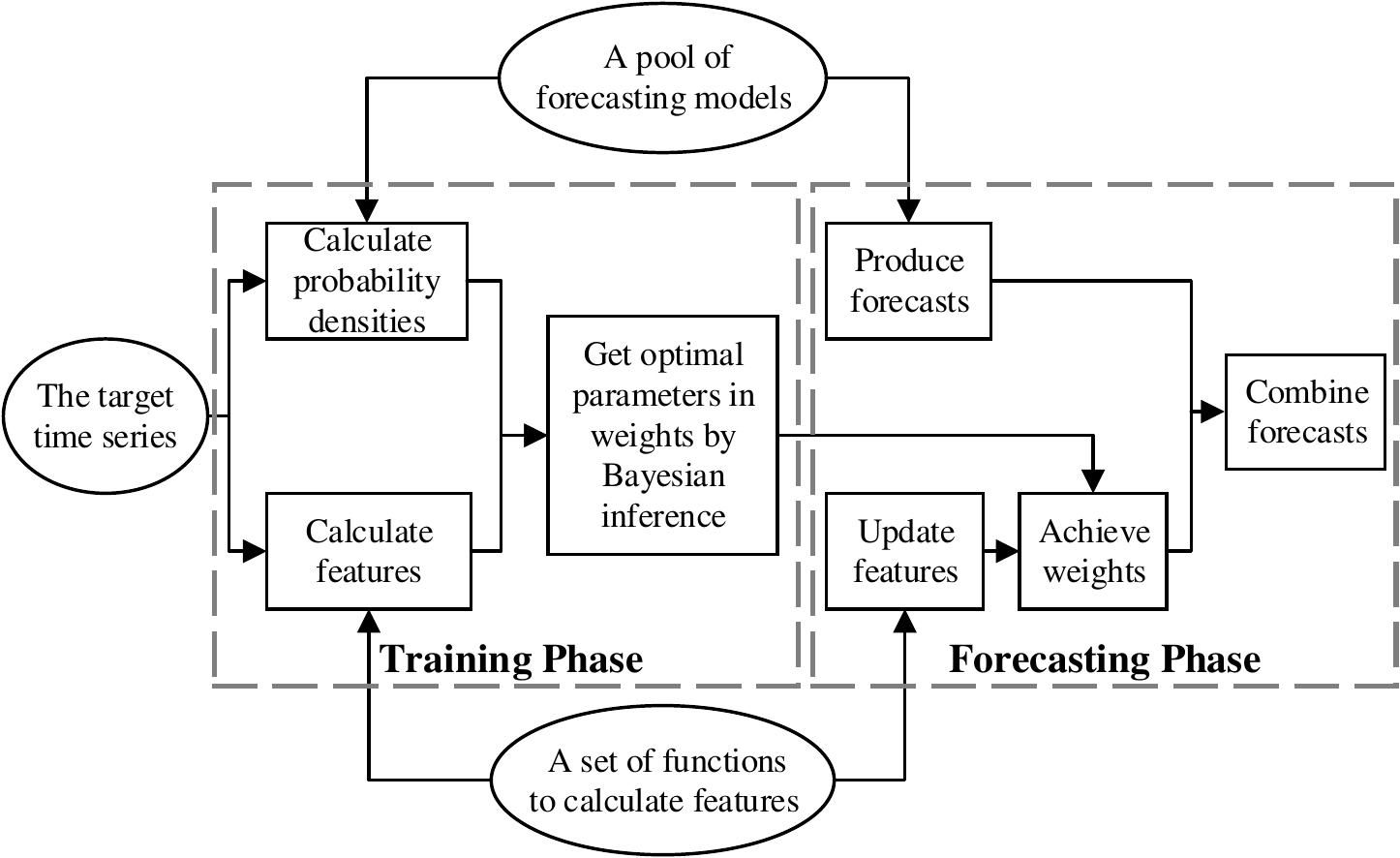}
    \end{tabular}
  \end{center}
  \caption
  { \label{framework}
    The implementation procedure for the proposed framework. }
\end{figure}

We put forward a two-phase procedure for the proposed framework as shown in Figure \ref{framework}. In the training phase, two parts of training data, that is, the time series features and probability predictive densities are required to get the optimal parameters. In the forecasting phase, we generate the combined forecast on the next phase based on the updated weights. This section provides the relevant details.

For the training phase, forecasters need to pick out relevant time series features and specify a pool of forecasting models. Then, probability predictive densities and features are calculated from observed values recursively. Thus, we obtain a matrix $P_{(T-s)\times m}$ within all the probability predictive densities of $m$ models and a matrix $F_{(T-s)\times n}$ consisting of $n$ features to train the optimal parameters in the time-varying weights. Although the computing time of this part is proportional to the length of the time series, the calculation is independent of different subsequences and can be implemented in parallel.

We utilize the \texttt{tsfeatures} R package \citep{Hyndman2019tsfeatures} to calculate features. Most of the features have been shown to perform well in prediction fields \citep{talagala2022fformpp,kang2019gratis,hyndman2015large-scale}. Then we apply the ReliefF algorithm \citep{kononenko1994estimating} to pick out relevant features. The number of features used in practice can be estimated by cross-validation (CV) to maximize the log score function.  Because the range of features differs, we standardize the calculated features to zero mean and variance of one to stabilize the optimization procedure. Furthermore, when the time series is long, it is time-consuming to calculate the granular features element-wisely. In this case, the computation of features can be carried out based on a sliding window of fixed length. The effect of the window size on the forecasting performance will be discussed in Section \ref{sec:stockdata}.

\begin{figure}
  \centering
  \includegraphics[width=\textwidth]{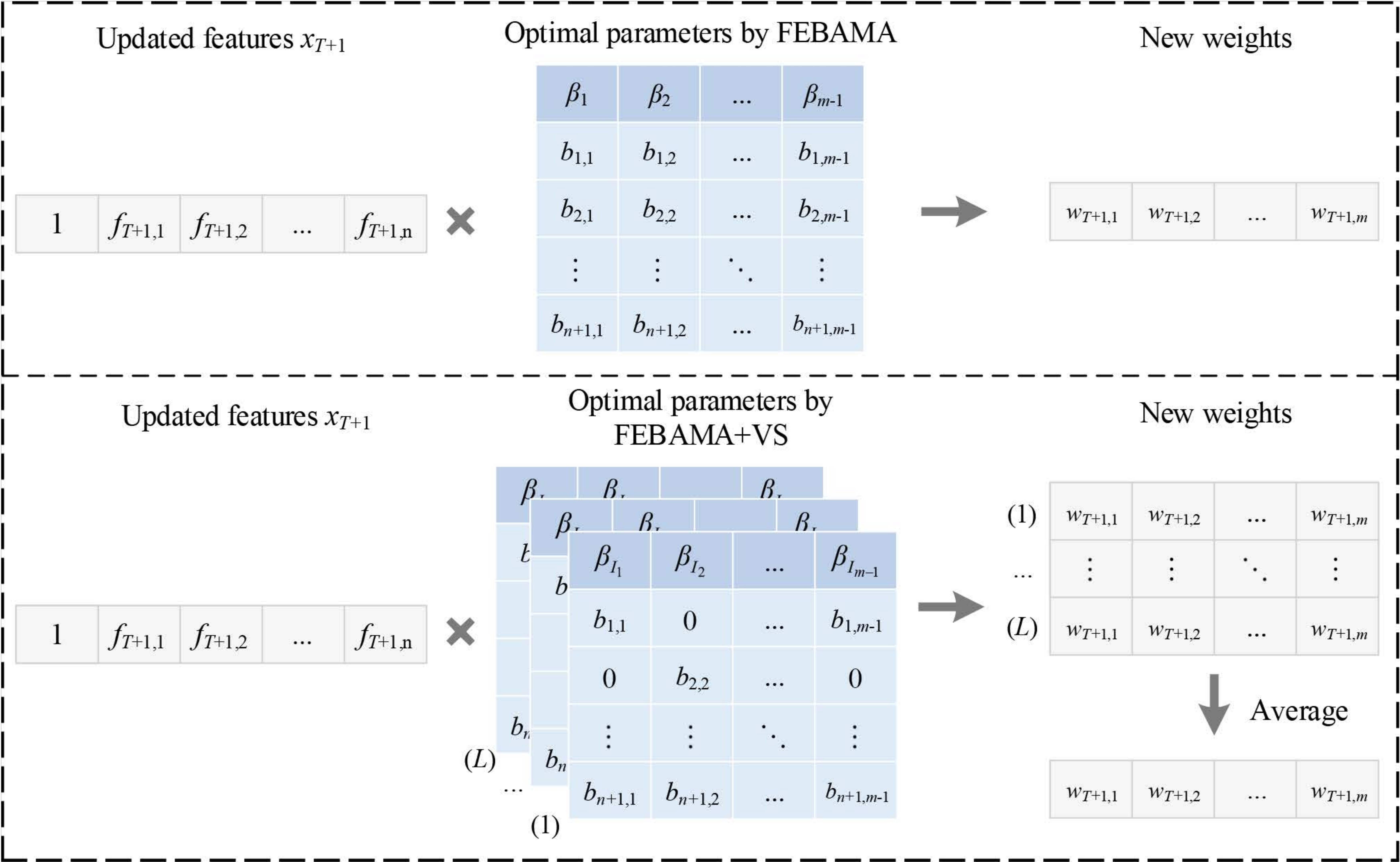}
  \caption{The weights for the one-step forecasting. The upper part shows the computation of weights based on an optimal coefficient matrix obtained by the FEBAMA method. The lower part presents the calculation of weights based on $L$ pairs of optimal coefficient matrices on the condition of variable selections in FEBAMA+VS method.}
  \label{forecast_weights}
\end{figure}

In the forecasting phase, we provide a linear combination of forecasts with updated weights. We use the optimal parameters from the training phase and recalculate features based on all historical data to generate the final weights in one-step forecast combination, as shown in Figure \ref{forecast_weights}.  For the FEBAMA method, $\left( {{\beta }_{1}},\cdots ,{{\beta }_{n-1}} \right)$ are the optimal coefficient vectors and the final weights can be directly obtained. Unlike original FEBAMA method that only one round of MAP is required, the FEBAMA+VS scheme runs $L$ pairs of $\left\langle\left( {{\mathcal{I}}_{\text{1}}},\cdots ,{{\mathcal{I}}_{n-1}} \right) ,\left( {{\beta }_{{\mathcal{I}}_{\text{1}}}},\cdots ,{{\beta }_{{\mathcal{I}}_{n-1}}} \right)\right\rangle$ via the Gibbs sampler to achieve a robust variable selection result. Thereafter, we calculate the mean of $L$ sets of weights as the final weights.

\begin{figure}
  \centering
  \includegraphics[width=0.8\textwidth]{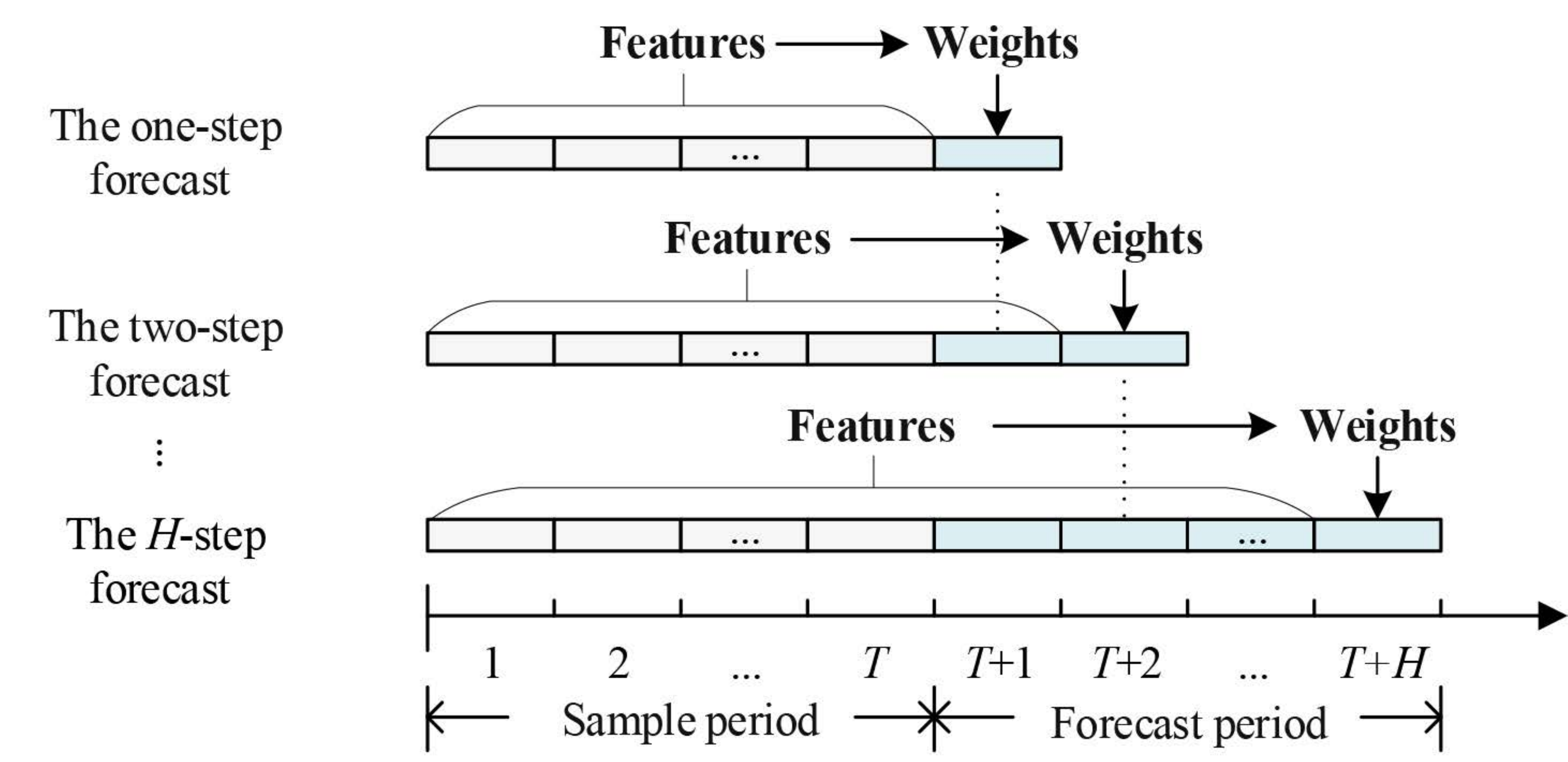}
  \caption{Scheme of one-step and multi-step forecasts. For example, in two-step forecasting, we calculate features based on a new time series by adding the one-step forecasted value to historical data. Then the final weights can be generated based on the updated features and the optimal parameters obtained by the proposed method.}
  \label{forecast}
\end{figure}

Once the weights in the forecast horizon are obtained, the one-step point forecast of the target series $Y_{T}$ can be combined as
\begin{align*}
  \label{eq:y_T+1_vs}
  {{\hat{y}}_{T+1}}=\sum\limits_{i=1}^{m}{{{{w}}_{T+1,i}}}{{\hat{y}}_{T+1,i}}.
\end{align*}
Additionally, the one-step density forecast is
\begin{equation*}
  \label{eq:pc_T+1_vs}
  p\left( {{{{y}}}_{T+1}};Y_{T} \right)=\sum\limits_{i=1}^{m}{{{{w}}_{T+1,i}}}p\left( {{{{y}}}_{T+1}}|Y_{T},{{M}_{i}} \right).
\end{equation*}
We produce multi-step forecasts recursively by regarding the one-step forecasted value as an observed value, as shown in Figure \ref{forecast}. The features are recalculated by including forecasted values to make the combination weights still time-varying inside the forecast period. We develop an R package \texttt{febama} available at \url{https://github.com/lily940703/febama}.

\section{Forecasting stock market data}
\label{sec:stockdata}

We illustrate the effectiveness and superiority of our framework through the following two experiments based on in-sample and out-of-sample forecasting, respectively. The simple average (SA), OP \citep{geweke2011optimal} and GP \citep{Kapetanios2015Generalised} are used as benchmark methods for comparison.

SA is a hard-to-beat forecast combination method, which simply combines forecasts with an equal weight ${{w}_{i}}=1/m$. \citet{clemen1989combining} reviewed over two hundred articles and concluded that the SA should be used as a benchmark. The phenomenon that SA performs better than more sophisticated combination methods is called the ``forecast combination puzzle'' in literature.  The puzzle is first stated in \citet{stock2004combination}, and its properties are studied empirically in later studies \citep{smith2009simple,claeskens2016forecast}.  Recently, \citet{Zhang2019Real-time} pointed out that combined methods with equal weights did not have worse forecast performance than those with time-varying weights.

OP \citep{geweke2011optimal} uses a series of predictive densities to construct optimal linear pools under the scoring criteria. The weights in density forecast combinations are obtained by maximizing the historical performance. Therefore, the optimal combination weights are constant. The OP is used as a benchmark in subsequent studies of density forecast combination with unsteady weights \citep{Kapetanios2015Generalised,Del2016Dynamic}. GP \citep{Kapetanios2015Generalised} extends OP to a more general scheme based on piecewise linear weight functions.  The weights in GP are estimated by maximizing the score of the generalized density combination and vary by region of the density.

\subsection{Data and problem description}

\begin{figure}
  \begin{center}
    \begin{tabular}{c}
      \includegraphics[width=\textwidth]{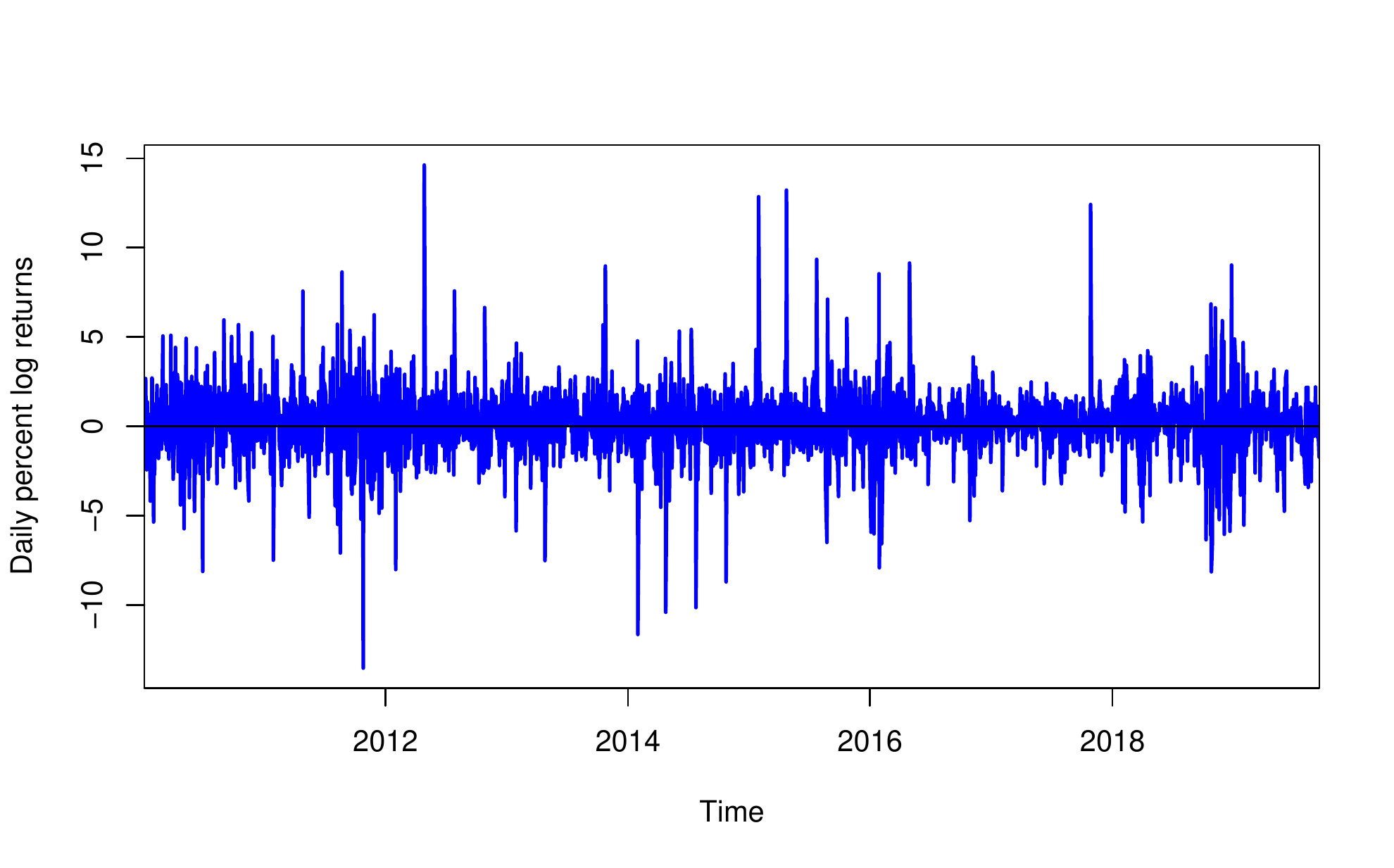}
    \end{tabular}
  \end{center}
  \caption
  { \label{fig:daiReturn}
    S\&P 500 daily percent log returns data from January 4, 2010 to  September 18, 2019.}
\end{figure}

The experiment considers daily percent log returns from January 4, 2010, to September 18, 2019, in the Standard and Poors (S\&P) 500 index, as shown in Figure \ref{fig:daiReturn}.  All the individual forecasting models are estimated by rolling samples of 1250 trading days (about five years). Then, we evaluate the proposed framework from two empirical studies. Firstly, we produce one-step density forecasts from December 19, 2014 ($t=1$) to September 18, 2019 ($t=1193$). We treat the 1,193 observations as an in-sample dataset. The combination weights in OP, GP and our proposed framework are optimized just once. In the second empirical study, we form out-of-sample density forecast combinations based on SA, OP, GP, and our proposed framework from $t=1$ to $t=1193$ by determining weights recursively, where only historical data are available. The two studies are based on one-step density forecasts, following the related empirical analysis in \citet{geweke2011optimal} and \citet{Kapetanios2015Generalised}. We refer to the parameter settings for priors in Table \ref{tab: priors} and the variance ${\sigma}^2$ is set to $10^3$. We measure the forecasting performance of density forecasts by LS, which is averaged over all the forecast values of the given series.

\subsection{Forecasting models and time series features}
\label{model_features}

We consider three basic forecasting models for the financial returns data: the Gaussian GARCH(1,1) model, the realized GARCH(1,1) model \citep{hansen2012realized} and the Stochastic Volatility (SV) model~\citep{Kastner2014Ancillarity}, denoted as $M_{1}$, $M_{2}$ and ${M}_{3}$, respectively.  The Realized GARCH (RGARCH) \citep{hansen2012realized} can jointly model the returns and the realized volatility, which is represented by the realized variance in this paper.  The first two models ${M}_{1}$ and ${M}_{2}$ are estimated via the maximum of likelihood, which are carried out using the \proglang{R} package \texttt{rugarch} \citep{Ghalanos_2014rugarch}. The third model ${M}_{3}$ is implemented using \proglang{R} package \texttt{stochvol} \citep{Kastner2019stochvol}.  In the following experiments, we implement the combination strategy and show the forecasting performance of combining each two of the three models, as well as combining all the three models.

In this work, we only utilize the widely used forecasting models with their standard forms. A series of volatility models including realized measures have flourished recently, such as the Heterogeneous AutoRegressive (HAR) \citep{corsi2009simple}, multivariate High-frEquency-bAsed VolatilitY (HEAVY) \citep{noureldin2012multivariate} and realized GARCH-Ito models \citep{song2021volatility}. Our focus in this paper is on the feature-based Bayesian forecasting combination. These specific models can be equally considered in the forecasting pool with our framework.

The time series features used in this experiment are calculated by the R package \texttt{tsfeatures} \citep{Hyndman2019tsfeatures}.  Firstly, we compute all the 42 features through a sliding window of the in-sample period. Then we filter out the features that do not change over time, such as seasonal features, and finally 31 features are reserved.  The number of features affects the forecasting performance of our method, as overmuch features increase the difficulty of the parameter estimation. Therefore, we need to further pick out relevant features to differentiate between different models.  We construct a multi-class classification problem by labeling the model that performs the best and employ the ReliefF algorithm \citep{kononenko1994estimating} to rank the features.  Finally, some of the top features are selected for our FEBAMA method.  We investigate the effect of two factors to calculate features (the window size and the number of features) in the in-sample forecasting evaluation.  The number of features for out-of-sample forecasting can be estimated to maximize the log predictive score based on CV.

\subsection{Evaluation of the in-sample forecasting performance}

For in-sample forecasting, Table \ref{tab:performance1_one_in} presents the average LS of the three individual models and forecast combination methods. We consider all possible composites of the three forecasting models (for example, $M_{1, 2}$ denotes the combination of models ${M}_{1}$ and ${M}_{2}$). ``Total'' in the last line means the average results of the four combinations.  As shown in Table \ref{tab:performance1_one_in}, the proposed FEBAMA method yields the highest LS for all the combinations, illustrating that the time-varying weights obtained by our method better match the objective of maximizing LS. In other words, taking time series features into consideration can improve the in-sample forecasting performance.  Note that the weights obtained by the FEBAMA+VS method are the average result based on different features in multiple variable selections. FEBAMA+VS helps augment stability and avoid overfitting in out-of-sample forecasting but has no prominent advantage in in-sample forecasting performance. Therefore, FEBAMA+VS is not included in Table \ref{tab:performance1_one_in}.

\begin{table}
  \centering
  \caption{In-sample log score comparison of individual models and different forecast combination methods. We consider all the possible combinations of $M_1$, $M_2$ and $M_3$ (e.g., $M_{1, 2}$ means the combination of $M_1$ and $M_2$). ``Total'' means the average results of the four combinations. For each combination (row), the highest LS is marked in bold. The last column refers to the $p$-values of the two-sided DM tests against the best method of the three benchmarks among SA, OP, and GP. The $p$-values less than 0.05 are marked in bold.}
  \begin{tabular}{lccccc}
    \toprule
    Individual   & ${M}_{1}$: GARCH & ${M}_{2}$: RGARCH & ${M}_{3}$: SV  &  &\\
                 & -1.3395          & -1.2867           & -1.2911    &    &  \\
    \midrule
    Combination  & SA               & OP           & GP & FEBAMA
    & DM test $p$-value\\
    \midrule
    $M_{1, 2}$   & -1.3138  & -1.2885    &  -1.2771    & \textbf{-1.2752} & 0.1749\\
    $M_{1, 3}$   & -1.3153  & -1.2911    &  -1.2850    & \textbf{-1.2783} & \textbf{0.0326}\\
    $M_{2, 3}$   & -1.2890  & -1.2832    &  -1.2770     & \textbf{-1.2709} & 0.0821\\
    $M_{1, 2, 3}$ & -1.3065 & -1.2827    &  -1.2716   & \textbf{-1.2626} & \textbf{0.0075}\\
    \midrule
    Total          & -1.3062 & -1.2864   &  -1.2777   & \textbf{-1.2717} & \\
    \bottomrule
  \end{tabular}%
  \label{tab:performance1_one_in}%
\end{table}%

To test for the significance of score differences during the in-sample period, we implement the two-sided Diebold-Mariano (DM) tests \citep{harvey1997testing}, which also has been applied to comparing forecasting methods \citep{li2022improving}. We only investigate the best method of the three benchmarks (SA, OP, and GP) and the proposed method in DM tests. The null hypothesis is that the two methods have the same forecast accuracy. The $p$-values in the last column of Table \ref{tab:performance1_one_in} indicate that, in three of out four situations, our FEBAMA method outperforms the best benchmark at 90\% level of significance.

It is worth mentioning that in the FEBAMA approach, there are two critical factors for the feature calculation: the window size and the number of features. A sliding window with proper length is essential because a short window makes the features noisy and a long window makes features steady. The window size is also connected with the length of the time series.  In the application to S\&P 500 returns, it is unreasonable to use all historical data to calculate granular features, because the calculation is time-consuming.

\begin{figure}
  \centering
  \includegraphics[width=\textwidth]{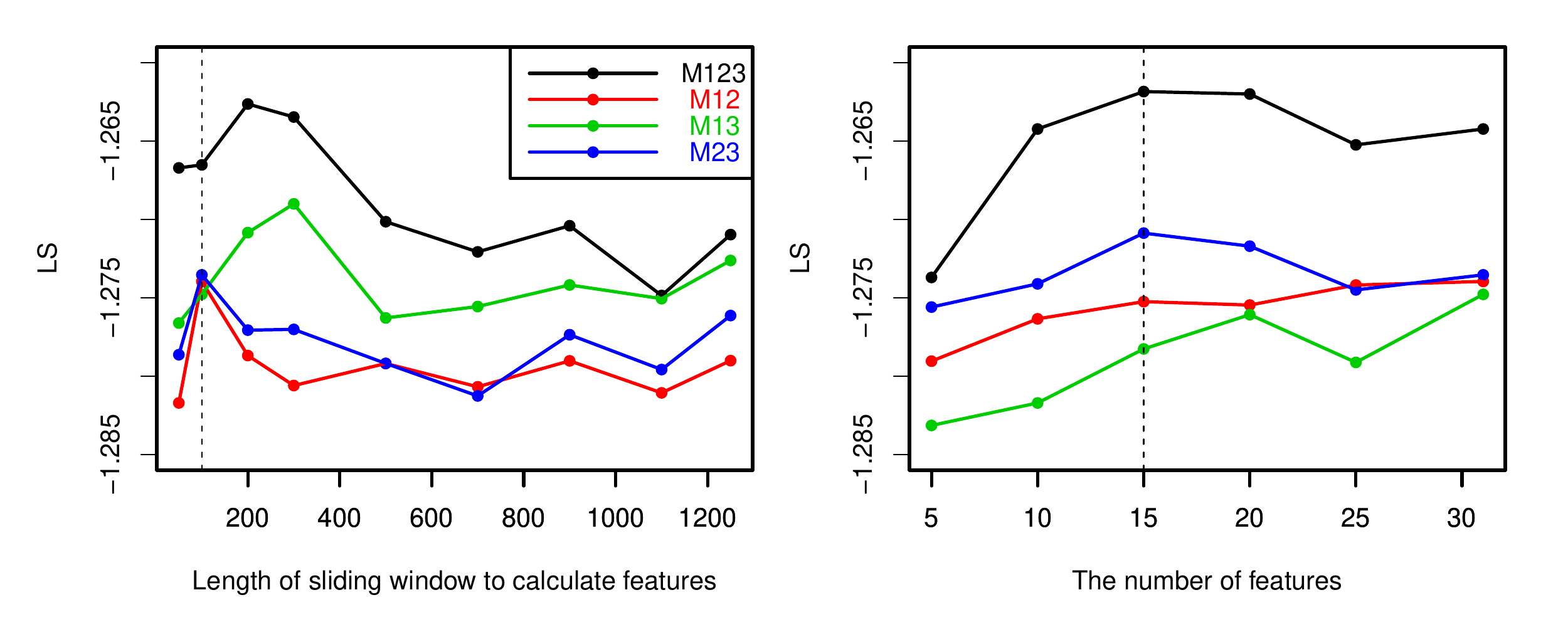}
  \caption{Plots of the effects of sliding window length and the number of features  on LS values.  We consider all the possible combinations of the three models (e.g., M12 means the combination of model $M_1$ and $M_2$).}
  \label{window_number}
\end{figure}

The left panel in Figure \ref{window_number} presents the change of the average LS with the length of the sliding window growing from 50 to 1,250 in this study, based on all four possible combinations. As the window size becomes larger, the forecast performance shows a trend of first rising and then declining.  We propose a simple principle for the window size selection --- selecting a smaller sliding window length so that the LS is as large as possible.  This strategy also simplifies the experiment and saves the calculation time. We choose the window length to be 100 for all the four combinations in Section~\ref{sec:stockdata}.

In addition, the number of features also plays an important role in the forecasting performance of our FEBAMA method. If the number of features is too small, the time series information that can be captured is restrictive. If the number is too large, the feature noise makes the algorithm difficult to converge. The right panel in Figure \ref{window_number} shows the in-sample forecasting results on LS with the number of features ranging from 5 to 31. With the increase of features, the forecast accuracy shows a general law of increasing first and then decreasing. We present the LS of the proposed FEBAMA based on 15 features in Table \ref{tab:performance1_one_in}. In the following experiment of Section \ref{out-of-sample}, we also use the 15 features, which are presented briefly in Table \ref{tab:features}. Some detailed descriptions can be found in previous work \citep{monteromanso2020fforma,kang2019gratis,talagala2022fformpp}.

\begin{table}%
  \centering
  \caption{Features used in forecasting stock market data.}
  \begin{tabular}{lp{9.5cm}l}
    \toprule
    Feature                           & Description                                                                           & Value range                     \\
    \midrule
    ${{F}_{\text{1}}}$: \textsf{alpha}        & ETS(A,A,N) $\widehat{\alpha}$                                                                      & $[0,1]$                     \\
    ${{F}_{\text{2}}}$: \textsf{arch\_acf}      & ARCH ACF statistic    & $\left( 0,\infty  \right)$ \\
    ${{F}_{\text{3}}}$: \textsf{arch\_r2}    & ARCH ${{R}^{2}}$ statistic   & $[0,1]$                    \\
    ${{F}_{\text{4}}}$: \textsf{beta}       & ETS(A,A,N) $\widehat{\beta}$    & $[0,1]$                     \\
    ${{F}_{\text{5}}}$: \textsf{crossing\_points}      & Number of times the time series crosses the median  & $\left\{ 1,2,3,\ldots  \right\}$                   \\

    ${{F}_{\text{6}}}$: \textsf{diff1x\_pacf5} & Sum of squares of first 5 PACF values of differenced series & $\left( 0,\infty  \right)$ \\
    ${{F}_{\text{7}}}$: \textsf{diff2\_acf1} & First ACF value of the twice-differenced series & $(-1,1)$\\
    ${{F}_{\text{8}}}$: \textsf{diff2\_acf10} & Sum of squares of first 10 ACF values of  twice-differenced  series & $\left( 0,\infty  \right)$\\
    ${{F}_{\text{9}}}$: \textsf{entropy} & Spectral entropy                                                                      & $(0,1)$                    \\
    ${{F}_{\text{10}}}$: \textsf{garch\_acf} & GARCH ACF statistic  & $\left( 0,\infty  \right)$ \\

    ${{F}_{\text{11}}}$: \textsf{garch\_r2} & GARCH ${{R}^{2}}$ statistic  & $[0,1]$                    \\
    ${{F}_{\text{12}}}$: \textsf{nonlinearity} & Nonlinearity coefficient &  $\left[ 0,\infty  \right)$\\
    ${{F}_{\text{13}}}$: \textsf{trend} & Strength of trend & $[0,1)$\\
    ${{F}_{\text{14}}}$: \textsf{unitroot\_kpss} & Test statistic based on KPSS test                                                     & $\left( 0,\infty  \right)$  \\
    ${{F}_{\text{15}}}$: \textsf{x\_acf1} &  First ACF value of the original series  &  $(-1,1)$\\
    \bottomrule
  \end{tabular}%
  \label{tab:features}
\end{table}%

\subsection{Evaluation of the out-of-sample forecasting performance}
\label{out-of-sample}

Table \ref{tab:performance1_one_out} shows the average LS of 1193 out-of-sample one-step forecasts based on different methods. We use predictive densities $p\left( {{y}_{k}};Y_{k-1},{{M}_{i}} \right),k=s+1,s+2,\cdots ,t$ to form the optimal predictive density for ${y}_{t+1}$ in OP, GP and our framework. As shown in Table \ref{tab:performance1_one_out}, the two methods based on the proposed framework are always the top two for all the possible combinations, which illustrates that the proposed framework improves the out-of-sample forecast accuracy. Furthermore, the FEBAMA+VS method achieves the largest LS all the time, because of its strengths regarding flexibility and stability.  The RGARCH and SV perform substantially better than GARCH, but the combination of the three models $M_{1, 2, 3}$ achieves a higher LS compared to the combination $M_{2, 3}$ based on FEBAMA+VS. An interesting finding is that a forecasting model that performs worse on its own can help improve forecasting when used in combination.

\begin{table}
  \centering
  \caption{Out-of-sample forecasting performance of the individual models and different forecast combination methods with regard to LS. We consider all the possible combinations of $M_1$, $M_2$ and $M_3$ (e.g., $M_{1, 2}$ means the combination of $M_1$ and $M_2$). ``Total'' means the average results of the four combinations. For each combination (in rowwise), the highest LS is marked in bold.}
  \begin{tabular}{lccccc}
    \toprule

    Individual & ${M}_{1}$: GARCH & ${M}_{2}$: RGARCH & ${M}_{3}$: SV &      &            \\
               & -1.3310          & -1.3065           & -1.3033            &              \\
    \midrule
    Combination    & SA       & OP    & GP       & FEBAMA        & FEBAMA+VS        \\
    \midrule
    $M_{1, 2}$       & -1.3109          & -1.3067     & -1.2984     & -1.2791       & \textbf{-1.2756} \\
    $M_{1, 3}$       & -1.3102          & -1.3054     &  -1.2913    & -1.2744       & \textbf{-1.2726} \\
    $M_{2, 3}$       & -1.3024          & -1.3008     &  -1.2958    & -1.2776       & \textbf{-1.2775} \\
    $M_{1, 2, 3}$       & -1.3031          & -1.3015  &  -1.2943       & -1.2768       & \textbf{-1.2703} \\
    Total            & -1.3067          & -1.3036     & -1.2950     & -1.2770       & \textbf{-1.2740} \\
    \bottomrule
  \end{tabular}%

  \label{tab:performance1_one_out}%
\end{table}%

\begin{figure}
  \centering
  \includegraphics[width=\textwidth]{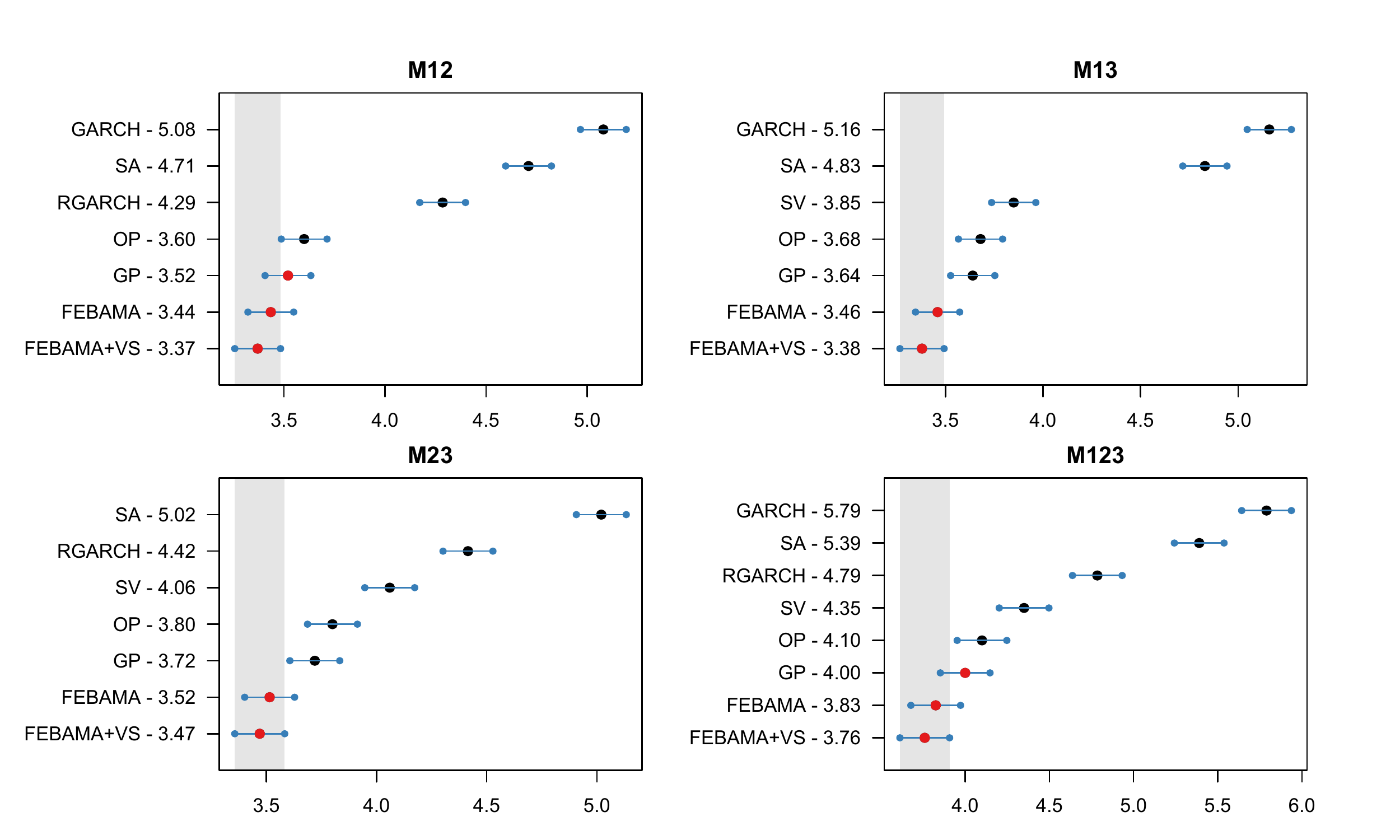}
  \caption{MCB test results for the ranks of different methods for all the possible combinations (e.g., M12 means the combination of model $M_1$ and $M_2$). The LS values assume a 95\% confidence level.}
  \label{fig:MCBtest}
\end{figure}

To identify whether the LS values of our FEBAMA approaches and other methods in Table \ref{tab:performance1_one_out} are statistically significant, we conduct Multiple Comparisons with the Best (MCB) \citep{koning2005m3} tests as shown in Figure \ref{fig:MCBtest}.  With MCB, Figure \ref{fig:MCBtest} indicates that the ranking performances of our methods are statistically better than others as a whole. Although FEBAMA+VS outperforms FEBAMA on average, their differences are not significant in terms of ranks. However, the variable selection can tell the contribution of different features, and the averaged weights achieve more stable forecasting results.

\begin{figure}
  \centering
  \includegraphics[width=\textwidth]{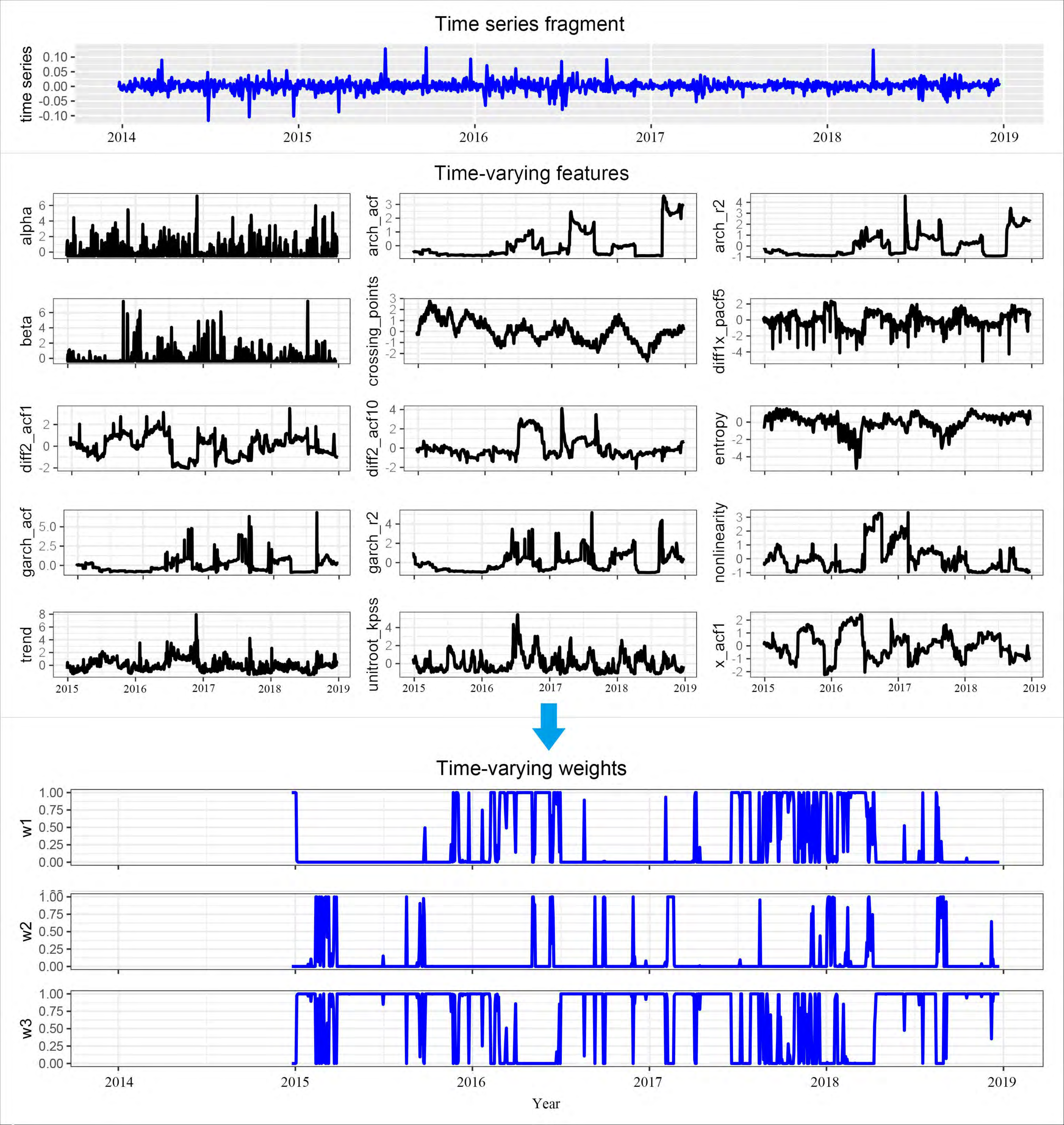}
  \caption{Time-varying features and combination weights for forecasting a time series fragment based on the proposed FEBAMA. The top panel shows the standardized time series. The 15 time-varying features are present in the middle panel, which are calculated based on historical data by a sliding window. The bottom panel shows time-varying weights obtained from time-varying features and the optimal parameters of FEBAMA in Table \ref{tab: coefficients}. The bottom panel denotes the weights of GARCH ($w_1$), RGARCH($w_3$), and SV($w_3$) models, respectively.}
  \label{fig:feat_weights}
\end{figure}

In the proposed framework, the time series features play a dominant role, not only improving the forecasting performance but also making the combination weights change with time and making it easy to explain. We show a time series fragment in Figure \ref{fig:feat_weights} to illustrate how the features affect the combination weights. Figure \ref{fig:feat_weights} presents the time-varying features (standardized) and the weights in combination $M_{1, 2, 3}$ based on a standardized time series fragment, whose length is 1,250. We set the minimum time series length to generate features to be 250. Based on FEBAMA, the estimated coefficient vectors in Equation \ref{eq:wit} are obtained in Table \ref{tab: coefficients}. Then the time-varying weights in the optimal forecast combination can be calculated through a linear function of features and a softmax transformation.

Furthermore, we can find a positive or negative correlation between the variation of time series features and combination weights in Figure \ref{fig:feat_weights}. For example, the crest of the feature ``\textsf{garch\_r2}'' leads to the peak of GARCH's weights. This is because the corresponding positive coefficient of 12.96 for ``\textsf{garch\_r2}'', and the absolute value of that is substantially larger than others in Table \ref{tab: coefficients} given all data are standardized.  As shown in Figure \ref{fig:feat_weights}, the stock prices rose significantly in 2015 and kept hitting record highs, unlike the continuous decline in 2014. The time series features in 2015 show up as high ``\textsf{entropy}'' values and low statistics related to ARCH and GARCH effects. These features lead to the high weights of the SV model in the stage of complex volatility.

Therefore, the time-varying features benefit the forecast combination by capturing the target series' characteristics.  The advantages of our method in the interpretation of weights are reflected in two aspects: (1) the weight at each time point can be explained by relevant features calculated from the historical data; and (2) the changing character of the time-varying weights in the whole sample period can be interpreted by the change rule of the time-varying features.

\begin{table}
  \centering

  \caption{The  estimated coefficients of features for the time series fragment in Figure \ref{fig:feat_weights}.}
  \begin{tabular}{c rrrrrrrr}
    \toprule
    & Intercept & ${{F}_1}$ & ${{F}_2}$ & ${{F}_3}$ & ${{F}_4}$ & ${{F}_5}$ & ${{F}_6}$ & ${{F}_7}$ \\
    \midrule
    $M_1$ & -12.06 & 2.33 & -10.48 & -0.51 & 1.95 & -4.44 & -4.61 & 3.20 \\
    $M_2$  & -24.57 & 2.63 & -18.54 & 21.06 & -2.89 & 10.26 & -8.31 & -0.58    \\
    \midrule
    & ${{F}_8}$ & ${{F}_9}$ & ${{F}_{10}}$ & ${{F}_{11}}$ & ${{F}_{12}}$ & ${{F}_{13}}$ & ${{F}_{14}}$ & ${{F}_{15}}$ \\
    \midrule
    $M_1$ & -7.91 & -11.87 & 8.20 & 12.96 &  -11.19 & 1.72 & -2.86 &  -8.07\\
    $M_2$  & -12.71 &  0.98 &  8.95 & 9.23 & 3.29 & -0.80 & -7.43  &  6.61 \\
    \bottomrule
  \end{tabular}
  \label{tab: coefficients}
\end{table}

\begin{figure}
  \centering
  \includegraphics[width=\textwidth]{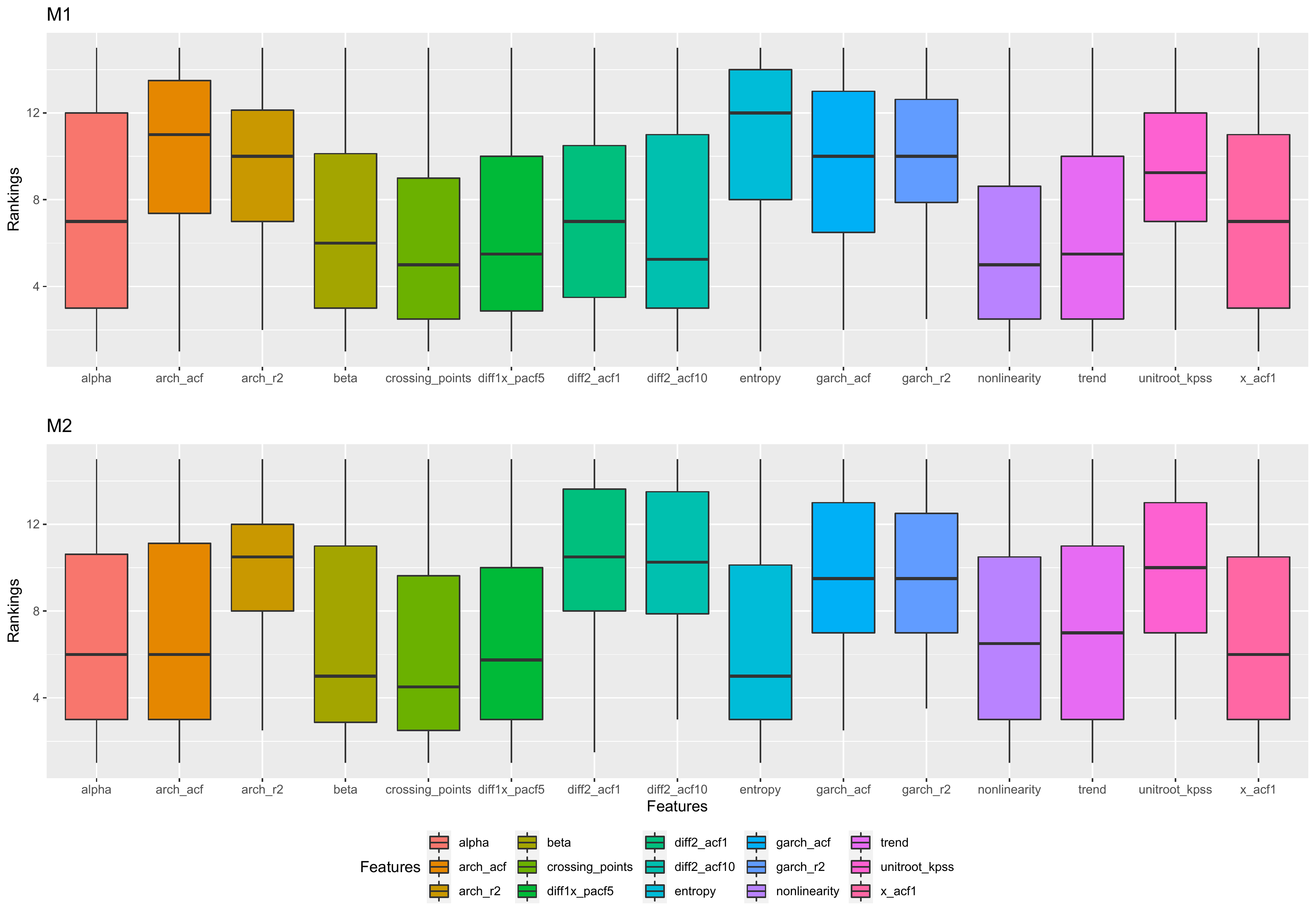}
  \caption{Boxplots of the rankings of the 15 features based on the combination $M_{1, 2, 3}$.  The ranking is  in ascending order, depending on the numbers that each feature is selected by FEBAMA+VS during the inference procedure. The top panel shows the results of the variable selection for model $M_1$, and the bottom is for model $M_2$.}
  \label{fig:ranks}
\end{figure}

The above analysis only focuses on the benefit of time series features in FEBAMA, but different features may contribute differently to the combining weights of the three forecasting models. FEBAMA+VS takes this into consideration, and thus delivers more stable results. Figure \ref{fig:ranks} shows the ranking distributions of the 15 features based on the combination $M_{1, 2, 3}$. The larger ranking indicates that the corresponding feature has been selected more frequently in the inference procedure and contributes more to the combination weights.  For the top model GARCH in Figure \ref{fig:ranks}, the ARCH/GARCH effects related features play an important role in the weights of $M_1$. ``\textsf{entropy}'', measuring the forecastability of time series, also matters a lot.  While for the bottom model RGARCH in Figure \ref{fig:ranks}, the features that measure the GARCH effect and the autocorrelation of twice-differenced series are most selected. ``\textsf{unitroot\_kpss}'' for the stationary test also has a significant effect in organizing the weights of $M_2$.  However, all 15 features contribute to the forecast combination as a whole. Our FEBAMA+VS method can conduct variable selections for forecasting models respectively and select different features for different models, leading to the prominence in flexibility.

\section{Forecasting the monthly M3 data}
\label{sec:m3data}

To study the performance of the proposed framework on other areas of data, we apply the proposed methods to M3 competition data~\citep{Makridakis2000M3}. The SA, OP \citep{geweke2011optimal} and GP \citep{Kapetanios2015Generalised} are also used as benchmark methods. Since the original OP and GP methods are designed for one-step forecasts, we use the constant weights estimated for one step to generate multi-step forecasts in the following experiments.

\subsection{Data and experimental setup}

The M3 data consist of 3,003 time series, covering yearly, quarterly, monthly, and other data from various areas such as micro, finance, and demographics industries and so on \citep{Makridakis2000M3}. We choose the monthly data in M3 in the current experiment, consisting of 1,428 series with lengths ranging from 48 to 126.  M3 competition required a forecasting horizon of 18 for monthly data \citep{Makridakis2000M3}. We thus carry out 18-step forecasts in the proposed framework, including point and density forecasts. We refer to the parameter settings for priors in Table \ref{tab: priors}, and the variance ${\sigma }^{2}$ is set to $10$. The forecasting performance measures are calculated by averaging all the horizons. Considering that the calculation of some features requires the time series length to be greater than two periods, the minimum length to generate prediction is 25 ($s=25$). Due to the limited time series length, all historical data are used to calculate features without considering a fixed-length sliding window for simplicity.

\subsection{Forecasting models and time series features used}

This experiment considers four individual forecasting models, which can be carried out using the \proglang{R} package \texttt{forecast} \citep{Hyndman2020forecast}. Relevant details are shown in Table \ref{tab:models}. The \proglang{R} package \texttt{forecast} provides the results of out-of-sample point forecasts and prediction intervals, which can be utilized to get probability predictive densities in our framework based on the normal distribution assumption.

\begin{table}%
  \centering
  \caption{Individual forecasting models for monthly M3 data.}
  \begin{tabular}{lll}
    \toprule
    Model             & Description                               & R implementation          \\
    \midrule
    $M_1$: \textsf{ets}        & Automated exponential smoothing algorithm & \texttt{ets(..., model = "AAN")} \\
    $M_2$: \textsf{naive}      & Naïve                                     & \texttt{naive()}                   \\
    $M_3$: \textsf{rw\_drift}  & Random walk with drift                    & \texttt{rwf(..., drift = TRUE)}  \\
    $M_4$: \textsf{auto.arima} & Automated ARIMA algorithm                 & \texttt{auto.arima()}              \\

    \bottomrule
  \end{tabular}%
  \label{tab:models}
\end{table}%

The features used in this experiment are calculated by the \texttt{tsfeatures} package \citep{Hyndman2019tsfeatures}. Like the previous experiment, we filter out the features that do not vary with time, and apply the ReliefF algorithm to select relevant features that can differentiate the forecasting performance of different models. Because of the diversity in the dataset, we pick out a different subset of the features to match each time series.  The number of features is estimated to be 6 by maximizing the average log score of the whole dataset.

\subsection{Forecasting performance for point and density forecasts}

We traverse all possible compounds of forecasting models, including a four-model combination $M_{1, 2, 3, 4}$, four three-model combinations [$M_{1, 2, 3}$, $M_{1, 2, 4}$, $M_{1, 3, 4}$, $M_{2, 3, 4}$], and six two-model combinations [$M_{1, 2}$, $M_{1, 3}$, $M_{1, 4}$, $M_{2, 3}$, $M_{2, 4}$, $M_{3, 4}$].  In this experiment, we compare our FEBAMA and FEBAMA+VS methods with three benchmarks SA, OP and GP for both point and density forecasts, results of which are listed in Table \ref{tab:performance_M3}.  Another popular feature-based forecast combination approach, FFORMA \citep{monteromanso2020fforma}, is however designed for point forecasting only.  We also performed FFORMA based on monthly M3 data with the MSAE measure, and presented the forecasting results in Table \ref{tab:performance_M3}.

From the left part of Table \ref{tab:performance_M3}, we can find that the average LS of SA, FEBAMA and FEBAMA+VS methods is consistently larger than that based on the individual models. The proposed FEBAMA+VS method is always in the top two and is the first in most combinations (9 out of 11).  Both exceptions occur in two-model combinations, which indicates that our method has more obvious strengths in complex multi-model combinations.  However, OP and GP are inapplicable in this experiment, as they do not beat SA in most cases, leading to the so-called ``forecast combination puzzle'' \citep{jeong2009combining}. As shown in Table \ref{tab:performance_M3}, $M_{1, 3, 4}$ is the preferred density forecasting combination with the maximum LS, although the component model ${{M}_{4}}$ is the worst individually.

\begin{table}
  \footnotesize
  \centering
  \caption{Average LS and MASE of different methods for density forecasts and point forecasts, respectively. The results for the individual models and all the possible combinations are presented. ``Total'' means the average results of all the combinations. For each combination (row), the highest LS and the lowest MASE are marked in bold.}
  \label{tab:performance_M3}
  \resizebox{\textwidth}{!}{
    \begin{tabular}{l ccccc cccccc}
      \toprule
      & \multicolumn{5}{c}{LS} & \multicolumn{6}{c}{MASE}                                                                                          \\
      \cmidrule(lr){2-6} \cmidrule(lr){7-12}
      Individual       & ${{M}_{1}}$            & ${{M}_{2}}$ & ${{M}_{3}}$    & ${{M}_{4}}$    &
                                                                               & ${{M}_{1}}$            & ${{M}_{2}}$ & ${{M}_{3}}$    & ${{M}_{4}}$    &                &                                                  \\
      & -4.368                 & -4.435      & -3.698         & -5.126         &                & 2.514 & 2.599 & 2.233 & 2.192  &        &        \\
      \midrule
      Combination      & SA                     & OP          & GP             & FEBAMA         & FEBAMA         & SA    & OP    & GP    & FFORMA & FEBAMA & FEBAMA \\
      &                        &             &                &                & +VS            &       &       &       &        &        & +VS    \\
      \cmidrule(lr){2-6} \cmidrule(lr){7-12}
      $M_{1, 2}$       & \textbf{-4.349}        & -4.372      & -4.368         & -4.358         & -4.352
                                                                               & 2.520                  & 2.513       & 2.514          & \textbf{2.512} & 2.519          & 2.514                                            \\
      $M_{1, 3}$       & -3.397                 & -3.283      & -3.265         & -3.224         & \textbf{-3.155}
                                                                               & 2.238                  & 2.207       & 2.212          & 2.208          & 2.219          & \textbf{2.193}                                   \\
      $M_{1, 4}$       & -3.524                 & -3.565      & -3.438         & -3.353         & \textbf{-3.282}
                                                                               & 2.228                  & 2.218       & 2.220          & 2.197          & 2.189          & \textbf{2.168}                                   \\
      $M_{2, 3}$       & -3.480                 & -3.280      & -3.257         & -3.255         & \textbf{-3.153}
                                                                               & 2.322                  & 2.257       & \textbf{2.243} & 2.262          & 2.277          & 2.245                                            \\
      $M_{2, 4}$       & -3.510                 & -3.527      & -3.432         & -3.331         & \textbf{-3.269}
                                                                               & 2.234                  & 2.208       & 2.218          & 2.190          & 2.195          & \textbf{2.172}                                   \\
      $M_{3, 4}$       & \textbf{-3.406}        & -3.842      & -3.814         & -3.665         & -3.611
                                                                               & 2.174                  & 2.135       & 2.146          & 2.154          & 2.123          & \textbf{2.108}                                   \\
      $M_{1, 2, 3}$    & -3.669                 & -3.273      & -3.561         & -3.187         & \textbf{-3.165}
                                                                               & 2.325                  & 2.208       & 2.263          & 2.243          & \textbf{2.138} & 2.201                                            \\
      $M_{1, 2, 4}$    & -3.754                 & -3.517      & -3.632         & -3.384         & \textbf{-3.252}
                                                                               & 2.307                  & 2.217       & 2.202          & 2.212          & 2.208          & \textbf{2.167}                                   \\
      $M_{1, 3, 4}$    & -3.158                 & -3.423      & -3.318         & -3.067         & \textbf{-3.064}
                                                                               & 2.135                  & 2.172       & 2.197          & 2.116          & 2.113          & \textbf{2.104}                                   \\
      $M_{2, 3, 4}$    & -3.178                 & -3.401      & -3.397         & -3.150         & \textbf{-3.098}
                                                                               & 2.147                  & 2.158       & 2.132          & 2.130          & 2.130          & \textbf{2.118}                                   \\
      $M_{1, 2, 3, 4}$ & -3.399                 & -3.480      & -3.523         & -3.178         & \textbf{-3.122}
                                                                               & 2.209                  & 2.172       & 2.221          & 2.165          & 2.153          & \textbf{2.120}                                   \\
      Total            & -3.529                 & -3.542      & -3.546         & -3.377         & \textbf{-3.320}
                                                                               & 2.249                  & 2.224       & 2.233          & 2.217          & 2.206          & \textbf{2.192}                                   \\
      \bottomrule
    \end{tabular}
  }
\end{table}

The mean absolute scaled error (MASE) \citep{Hyndman2006Another} is a forecasting accuracy metric used in point forecasts. The MASE is scaled by the mean absolute error of in-sample one-step forecasts using the naive forecasting method as follows

\begin{equation}
  \label{eq:mase}
    MASE=\frac{\sum\nolimits_{h=1}^{H}{\left| {{\hat{y}}_{T+h}}-{{y}_{T+h}} \right|/H}}{\sum\nolimits_{t=2}^{T}{\left| {{y}_{t}}-{{y}_{t-1}} \right|/\left( T-1 \right)}}.
\end{equation}

${{\hat{y}}_{T+h}}$ is point forecast, and $h$ is the forecasting horizon, $h=1,2,...,H$. \citet{Hyndman2006Another} recommended it to be the standard measure for forecast accuracy because of its excellent properties such as scale independence and easy interpretability.

We utilize MASE to measure the performance of point forecasts, where lower MASE is better. The MASE panel of Table \ref{tab:performance_M3} shows that the proposed method outperforms others in a majority of combinations, except for $M_{1, 2}$ and $M_{2, 3}$. But FEBAMA+VS presents a similar performance to the best approach for the two combinations.  Although the target of the proposed framework is to maximize LS based on predictive densities, our method still achieves outstanding performance in point forecasts, which further illustrates the prominence and flexibility of the proposed framework.

In the last row of Table \ref{tab:performance_M3}, we summarize the average performance of different methods based on all model combinations.  The results show that the two methods in our framework are significantly superior to the three benchmarks in both LS and MASE.  In contrast to the last experiment in Section \ref{sec:stockdata}, the experiment based on the monthly data in the M3 competition demonstrates the advantages of the proposed FEBAMA framework from the following three novel perspectives.

\begin{itemize}
\item The performance of point forecasts can also be improved based on the combination weights estimated from FEBAMA, which was designed to maximize LS for density forecasts.

\item  When historical data are limited, OP and GP are challenging to estimate the optimal weights in the density forecast combination to get better results over SA. The proposed FEBAMA framework still works well by learning the relationship between time-varying features and combination weights.

\item For multi-step forecasting, we obtain the time-varying weights in the forecast horizon by adding predicted values to calculate time-varying features. The performance that is based on 18-step forecasts in Table \ref{tab:performance_M3} shows the effectiveness and superiority of FEBAMA in the field of multi-step forecasting.
\end{itemize}

To test for the significance of performance differences in Table \ref{tab:performance_M3}, we also implement the two-sided DM tests \citep{harvey1997testing} and show the $p$-values in Table \ref{tab:DMtest}. We only output the $p$-values of DM tests between our proposed methods and the best method of the three benchmarks (SA, OP, and GP) according to Table \ref{tab:performance_M3}. The results in Table \ref{tab:DMtest} indicate that three and four cases of the FEBAMA methods, in terms of  LS and MASE respectively, are better than the best benchmark among SA, OP, and GP methods over all 11 possible combinations at 95\% level of significance.  The significant cases, which are also dominated by the multiple combinations, increase to seven with the FEBAMA+VS, indicating the benefit of variable selection.

\begin{table}
  \footnotesize
  \centering
  \caption{The $p$-values of the two-sided DM tests for comparing the forecasting accuracy of the best benchmark among SA, OP, and GP in Table \ref{tab:performance_M3}  with the proposed methods. The null hypothesis is that the two approaches have the same forecast accuracy. The $p$-values less than 0.05 are marked in bold.}
  \label{tab:DMtest}%
  \resizebox{\textwidth}{!}{
    \begin{tabular}{l ccccc ccccc c}
      \toprule
      &\multicolumn{11}{c}{LS}                                            \\
      \cmidrule(lr){2-12}
      & $M_{1, 2}$ & $M_{1, 3}$ & $M_{1, 4}$ & $M_{2, 3}$ & $M_{2, 4}$ & $M_{3, 4}$ & $M_{1, 2, 3}$ & $M_{1, 2, 4}$  & $M_{1, 3, 4}$ & $M_{2, 3, 4}$ & $M_{1, 2, 3, 4}$ \\
      FEBAMA      & 0.612 & 0.183 & 0.360 & 0.152& 0.232& 0.673& 0.153 & \textbf{0.032} & 0.232 & \textbf{0.028} & \textbf{0.038}  \\
      {FEBAMA+VS} & 0.535 & \textbf{0.045} & \textbf{0.037} & \textbf{0.006} & 0.067 & 0.642& 0.249 & \textbf{0.023} & \textbf{0.004} & \textbf{0.020} &  \textbf{0.034} \\

      \midrule
      &\multicolumn{11}{c}{MASE}  \\
      \cmidrule(lr){2-12}
      & $M_{1, 2}$ & $M_{1, 3}$ & $M_{1, 4}$ & $M_{2, 3}$ & $M_{2, 4}$ & $M_{3, 4}$ & $M_{1, 2, 3}$ & $M_{1, 2, 4}$  & $M_{1, 3, 4}$ & $M_{2, 3, 4}$ & $M_{1, 2, 3, 4}$ \\
      FEBAMA      & 0.733 & 0.218 & \textbf{0.022} & 0.542 & 0.196 & \textbf{0.046} & \textbf{0.023} & 0.259 & 0.079 & \textbf{0.017} &  0.216\\
      {FEBAMA+VS} & 0.377 & 0.220 & \textbf{0.018} & 0.338 & \textbf{0.050} & \textbf{0.022} & \textbf{0.006} & \textbf{0.018} & \textbf{0.000} & \textbf{0.004} &  0.242\\
      \bottomrule
    \end{tabular}%
  }
\end{table}%

\section{Discussion}
\label{sec:discussion}

Density forecast combinations with fixed and time-varying weights have achieved growing attention recently. Our proposed framework enriches this vein by obtaining the time-varying weights based on aggregating time-varying features. To the best of our knowledge, this is the first time features are taken into consideration for time-varying forecast combinations. In contrast to black-box forecasting combination schemes, our proposed framework has the significance for interpretability: (1) the time-varying weights at each time point can be expressed by time series features calculated from historical data; (2) the variation of time-varying weights can be explained by the trend of related time-varying features; and (3) the contribution of different features to the forecast combination can be measured through an automatic Bayesian variable selection method with the proposed framework.  We summarize the characteristics of our approaches in Table \ref{tab:Comparison} based on a comparison with the SA, OP, GP, and FFORMA methods.

\begin{table}
  \centering
  \caption{Comparison between different combination methods.}
  \resizebox{\textwidth}{!}{
  \begin{tabular}{lcccccc}
    \toprule
    & {SA} & {OP} & GP & FFORMA & FEBAMA & FEBAMA \\
    &&&&&&+VS\\
    \midrule
    The weights are optimized                                & N    & Y    & Y       &      Y         & Y        & Y           \\
    The weights are time-varying                             & N    & N    & Y       &       N        & Y        & Y           \\
    Take features into consideration                         & N    & N    & N       &        Y       & Y        & Y           \\
    Measure the importance of different time series features & N    & N    & N       &        N       & N        & Y           \\
    \bottomrule
  \end{tabular}%
  }
  \label{tab:Comparison}%
\end{table}%

Our framework shows great superiority in both S\&P 500 and M3 competition data, indicating that our method is applicable to data of various lengths and from diversified fields. Furthermore, the proposed framework trains optimal parameters based on simple statistical features and the historic forecasting performance, with no need for mass data to learn the relationship between weights and features in contrast to some machine learning methods \citep{monteromanso2020fforma,li2020forecasting,wang2021uncertainty}.

Computational efficiency also needs attention, especially when the data set is large or the historical data are long. Because the forecasting of each time series is independent, the proposed framework is parallelizable. Particularly, the training period of our framework is often the most time-consuming part, on account of the calculation of predicted densities and features. We can complete the process in the offline phase. In addition, we further shorten the computing time by using rolling samples in forecasting returns data.

Although we formulated a full Bayesian scheme for time-varying forecasting combination, the inference of the coefficients of time series features is based on a simple MAP scheme and the variable selection process is also simple. Variational inference and stochastic gradient based methods \citep{2011Bayesian,2014Stochastic} could be further explored. The proposed framework is designed for density combinations, but potential users may have forecasts only available in the form of a sample from the predictive distribution.  In these circumstances, the variance of the forecast error needs to be estimated through techniques such as bootstrap or ensemble learning to form an empirical distribution. Our framework could be extended to more application scenarios.

Furthermore, the objective of our framework is to obtain the optimal weights in forecast combinations.  Current work has neglected to select an appropriate pool of forecasting models, known as trimmed linear pooling \citep{grushka2017ensembles} or forecast pooling \citep{Kourentzes2019Another}. In the two experiments of Section~\ref{sec:stockdata} and Section \ref{sec:m3data}, we traverse all possible composites of forecasting models and find some interesting results. For instance, combining all models is not the best strategy. The improvement of some combinations is significant compared to individual models, but the advantages of other combinations are not obvious. Therefore, constructing an appropriate forecast pool before the proposed framework is essential, especially with abundant alternative models.

\section{Conclusions}
\label{sec:conclusions}

Forecast combinations can tackle model uncertainty in forecasting and bring performance improvements. The proposed framework extends existing density forecast combinations from an innovative proposition, allowing the combination weights to vary with time-varying features. Specifically, we apply an automatic Bayesian variable selection method to identify the importance of different features. Therefore, our framework can interpret which features and the contribution of features to determine the combination weights of each model, in contrast to black-box combination methods.

To emphasize the strengths of the proposed framework, we use benchmark combination schemes with both fixed weights and time-varying weights.  Two experimental studies based on S\&P500 returns and M3 competition data show that the proposed methods can produce more accurate point and density forecasts for one step or multiple steps than the benchmark methods. Factors that may affect the forecasting performance are also discussed, including different combinations of individual models, the number of selected features and the window size for feature calculation.

\section*{Declaration of competing interest}

The authors declare that they have no known competing financial interests or personal relationships that could have appeared to influence the work reported in this paper.

\section*{Acknowledgments}

The authors are grateful to the editors and three anonymous reviewers for helpful comments that improved the contents of the paper.


\printbibliography
\end{document}